\documentclass[pre,aps,twocolumn,superscriptaddress,floatfix]{revtex4-1}
\usepackage[english]{babel}
\usepackage[dvipsnames]{xcolor}
\usepackage{amsmath} 
\usepackage{color}
\usepackage{graphicx}
\usepackage{dcolumn}
\usepackage{bm}
\usepackage[colorlinks=true, linkcolor=blue, citecolor=blue, urlcolor=blue]{hyperref}
\usepackage{float} 

\begin{document}



\title{Roughening of active nonlinear interfaces with broken 
tilt symmetry}

\author{A. M. Cámara}
\affiliation{Instituto de Investigaciones Físicas de Mar del Plata (IFIMAR), Facultad de Ciencias Exactas y Naturales, Universidad Nacional de Mar
del Plata, Consejo Nacional de Investigaciones Científicas y Técnicas (CONICET), Deán Funes 3350, B7602AYL Mar del Plata, Argentina}

\author{A. B. Kolton}
\affiliation{Centro At\'omico Bariloche, CNEA, CONICET, Bariloche, Argentina}
\affiliation{ Instituto Balseiro, Universidad Nacional de Cuyo, Bariloche, Argentina}

\author{J. L. Iguain}
\affiliation{Instituto de Investigaciones Físicas de Mar del Plata (IFIMAR), Facultad de Ciencias Exactas y Naturales, Universidad Nacional de Mar
del Plata, Consejo Nacional de Investigaciones Científicas y Técnicas (CONICET), Deán Funes 3350, B7602AYL Mar del Plata, Argentina}

\date{\today}
\begin{abstract}
We study the roughening of an interface with nonlinear elasticity driven by temporally correlated noise, which breaks statistical tilt symmetry. Using scaling arguments and a self-consistent Hartree approximation, we derive the crossover diagram and the steady-state structure factor. We identify three scaling regimes associated with the Larkin, anharmonic Larkin, and Edwards--Wilkinson universality classes, and obtain the crossover lengths separating them. Numerical simulations of large systems confirm the analytical predictions over the full parameter range. Our results provide a unified description of finite-size and crossover effects in a minimal nonlinear-elastic Ornstein--Uhlenbeck active interface.
\end{abstract}

\pacs{Valid PACS appear here}

\maketitle

\section{Introduction and Model}
\label{sec:intro}

Interfaces driven far from equilibrium play a central role in a wide range of physical systems, from crystal growth~\cite{BarabasiStanley1995} and field-driven magnetic domain walls~\cite{lemerle1998domain, metaxas2007creep, gorchon2014pinning} to reaction fronts~\cite{KoltonLaneri2019,zagarra2024}, biological membranes~\cite{Safran1994,ChaikinLubensky1995,NelsonPiranWeinberg2004}, and active materials~\cite{marchetti2013hydrodynamics, matsushita1998interface, sussman2018soft, bru2003universal, galeano2003dynamical}. A central challenge is to understand how fluctuations and elasticity determine not only their large-scale roughening properties but also the finite-size effects and scale-dependent crossovers that arise away from asymptotic scaling, as recently emphasized in active matter systems~\cite{Gompper_2025}.

The Edwards--Wilkinson (EW) and Kardar--Parisi--Zhang (KPZ) equations~\cite{BarabasiStanley1995} provide the standard paradigms for kinetic roughening and universality in growing interfaces. However, their asymptotic scaling descriptions typically assume white-noise driving and harmonic elasticity, and therefore do not capture the crossover phenomena generated by persistent forcing and nonlinear elastic responses. In many physical systems however, the driving forces acting on an interface are temporally correlated, reflecting the persistence of microscopic active components or environmental fluctuations~\cite{Marchetti2013,Fodor2016}. Such colored noise generically leads to scale-dependent crossover behavior and violates the assumptions underlying equilibrium descriptions based on white-noise driving of hamiltonian models.
Concurrently, the mechanical response of physical interfaces is often anharmonic, particularly under large local distortions or when microscopic structure induces non-linear restoring forces~\cite{ChaikinLubensky1995}. Nonlinear elasticity can significantly modify scaling behavior and is not captured by harmonic interface models.

Capturing both effects---colored noise and nonlinear elasticity---within a unified continuum description is therefore essential to bridge the gap between idealized asymptotic theories and experimentally accessible realizations.
We therefore consider a minimal continuum model that incorporates both temporally correlated driving and nonlinear elasticity in a controlled setting. Specifically, we focus on an overdamped interface with gradient-flow dynamics derived from a local elastic energy and driven by an Ornstein--Uhlenbeck colored noise process. This choice preserves the equilibrium structure in the white-noise limit while providing a simple framework to study nonequilibrium effects induced by finite noise persistence.

From a broader perspective, this setup belongs to the class of Ornstein–Uhlenbeck (OU) active matter models, whose statistical mechanics and thermodynamics have been extensively analyzed in the context of active particles with various self-propulsion styles and interactions~\cite{Bonilla2019, Martin2021}. By considering a time-independent friction response, our framework violates the equilibrium fluctuation–dissipation theorem (FDT) when the noise correlation time is finite, serving as a useful, minimal continuum model for exploring the non-equilibrium features of active elastic interfaces.

Specifically, we investigate the overdamped dynamics of a one-dimensional interface with anharmonic corrections to its elasticity, subject to a short-range temporally correlated but spatially uncorrelated active noise $\eta(x,t)$:
\begin{equation}
 \partial_t h(x,t) = c_2 \partial_x^2 h(x,t) + c_{2n} \partial_x \left[(\partial_x h(x,t))^{2n-1}\right] + \eta(x,t),
\label{eq:eqmotion}
\end{equation}
where $n>1$ is an integer determining the degree of non-Hookean elasticity, while $c_2>0$ and $c_{2n}>0$ represent the linear and non-linear elastic constants, respectively. The stochastic driving force $\eta(x,t)$ satisfies $\langle \eta(x,t) \rangle = 0$ and is exponentially correlated in time:
\begin{equation}
\langle \eta(x,t)\eta(x',t') \rangle = \frac{T}{\tau} \delta(x-x') e^{-\frac{|t-t'|}{\tau}},
\label{eq:colorednoise}
\end{equation}
characterized by the persistence time $\tau>0$ and a power spectrum given by:
\begin{equation}
S_\eta(k,\omega)= \frac{2T}{1+\omega^2 \tau^2}.
\label{eq:noisepowerspectrum}
\end{equation}
The colored noise profile can be generated via the steady-state solution of the auxiliary coupled Markovian equation:
\begin{equation}
\tau \partial_t \eta(x,t) = -{\eta(x,t)} + \xi(x,t),
\label{eq:noiseeqcont}
\end{equation}
where $\xi(x,t)$ is a standard Gaussian white noise with $\langle \xi(x,t) \xi(x',t') \rangle = 2T \delta(x-x') \delta(t - t')$. 

The non-linear term $\partial_x[(\partial_x h)^{2n-1}]$ derives directly from the total elastic energy functional:
\begin{equation}
E = \int \mathrm{d}x \left[\frac{c_2}{2} (\partial_x h)^2 + \frac{c_{2n}}{2n} (\partial_x h)^{2n} \right],
\end{equation}
such that the deterministic evolution can be written as an energy minimization gradient descent, $\partial_t h = -\delta E / \delta h + \eta(x,t)$. For any finite persistence time $\tau$, the system generates a non-equilibrium steady state. Notably, this dynamics preserves reflection symmetry ($h\to -h$) but fundamentally breaks statistical tilt symmetry (STS), $h(x,t) \to h(x,t) + s \cdot x$, under an imposed spatial slope $s$. These structural symmetries are exactly the inverse of the paradigmatic KPZ equation, which breaks $h \to -h$ but preserves STS. As a consequence of this broken symmetry, the non-linear coupling constant $c_{2n}$ is unprotected and renormalizes independently from $c_2$ under coarse-graining.

For any finite $\tau$, the violation of the FDT triggers genuine non-equilibrium interface dynamics accompanied by continuous entropy production~\cite{Cugliandolo_2011,Fodor2016}. In the limit $\tau \to 0$, the persistence time vanishes, the driving force maps onto an ordinary thermal white noise, and the system relaxes into an equilibrium Gibbs-Boltzmann steady state $P(h) \propto \exp(-E/T)$ where the FDT is strictly satisfied at temperature $T$. Conversely, in the frozen limit $\tau \to \infty$, the noise transforms into a static, quenched random force landscape $\eta(x)$. In this asymptotic regime, Eq.~\eqref{eq:eqmotion} reduces to the anharmonic Larkin model (ALM)~\cite{purrello2019roughening}, which exhibits an anomalous, multi-affine structural roughening characterized by two distinct super-roughness exponents: a global roughness exponent $\zeta = (4n - 1) / (4n - 2)$ (for $d=1$) and a larger spectral exponent $\zeta_s > \zeta$.

For finite, non-zero values of $\tau$ and $T$, the full steady-state behavior of Eq.~\eqref{eq:eqmotion} have yet to be uncovered. The core objective of this paper is to unveil this comprehensive crossover landscape and map the multi-scale roughening properties of the non-linear active interface as a function of the active persistence $\tau$ and the environmental fluctuations amplitude $T$, for competing linear and nonlinear elasticities.

\begin{figure}[t]
    \centering
    \includegraphics[width=\linewidth]{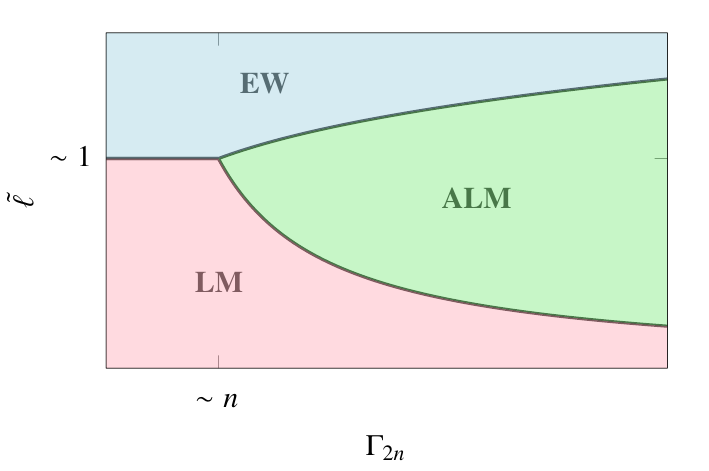}
    \caption{Crossover diagram for the non-linear active interface described by Eq.~\eqref{eq:eqmotion}, mapped as a function of the normalized length scale $\tilde{\ell} = \ell/\sqrt{c_2 \tau}$ and the  dimensionless coupling constant $\Gamma_{2n} = c_{2n} T / (c_2^{(3n-1)/2}\sqrt{\tau})$. Here, $T$ denotes the active noise strength, $\tau$ is the temporal correlation time, $c_2$ is the linear elastic constant, and $c_{2n}$ (with $n>1$) represents the degree of non-Hookean elasticity. The analytical boundaries mapping the transitions between the EW, LM, and ALM roughening regimes are defined by Eqs.~\eqref{eq:line1}--\eqref{eq:line3}.} 
    \label{fig:crossoverdiagram}
\end{figure}

\section{Summary of Main Results}
\label{sec:summary_results}

We find that the interplay between active noise persistence and nonlinear elasticity organizes the steady-state roughening dynamics into a sequence of three self-affine regimes: Larkin (LM), anharmonic Larkin (ALM), and Edwards--Wilkinson (EW). Each regime is characterized by distinct scaling exponents, which we determine analytically (see Table~\ref{tab:exponents3regimes}). These predictions are supported by large-scale numerical simulations.

The resulting behavior is summarized in the crossover diagram of Fig.~\ref{fig:crossoverdiagram}, which is expressed in terms of the normalized length scale 
\begin{equation}
 \tilde{\ell} = \ell / \sqrt{c_2 \tau}
\label{eq:adimensionallength}
\end{equation}
and the dimensionless coupling
\begin{equation}
\color{black}
\Gamma_{2n} = \frac{c_{2n} T}{c_2^{(3n-1)/2} \sqrt{\tau}}.
\color{black}
\label{eq:gamma_coupling_definition}
\end{equation}
This coupling quantifies the competition between nonlinear elasticity and linear restoring forces under temporally correlated driving.

The boundaries between scaling regimes are given by
\begin{align}
\tilde{\ell}_{\rm LM/EW} &\sim 1, 
\label{eq:line1}
\\
\tilde{\ell}_{\rm ALM/EW} &\sim (\Gamma_{2n}/n)^{1/(3n-1)}, 
\label{eq:line2}
\\
\tilde{\ell}_{\rm LM/ALM} &\sim 
\color{black}
(\Gamma_{2n}/n)^{-1/(n-1)},
\color{black}
\label{eq:line3}
\end{align}
so at a given $\Gamma_{2n}$, there are two crossovers if $\Gamma_{2n} > n$ and only one crossover otherwise.

While the EW and LM regimes correspond to standard Edwards--Wilkinson and Larkin universality classes, the emergent intermediate ALM regime exhibits anomalous super-rough behavior induced by nonlinear elasticity and the breaking of statistical tilt symmetry. This regime corresponds to the extension of the faceted roughening class known from the quenched (``frozen'') limit $\tau \to \infty$~\cite{purrello2019roughening}.

\section{Linear Active Interface}
\label{sec:linearlimit}

\begin{figure}[t]
\centering
\includegraphics[width=1\columnwidth]{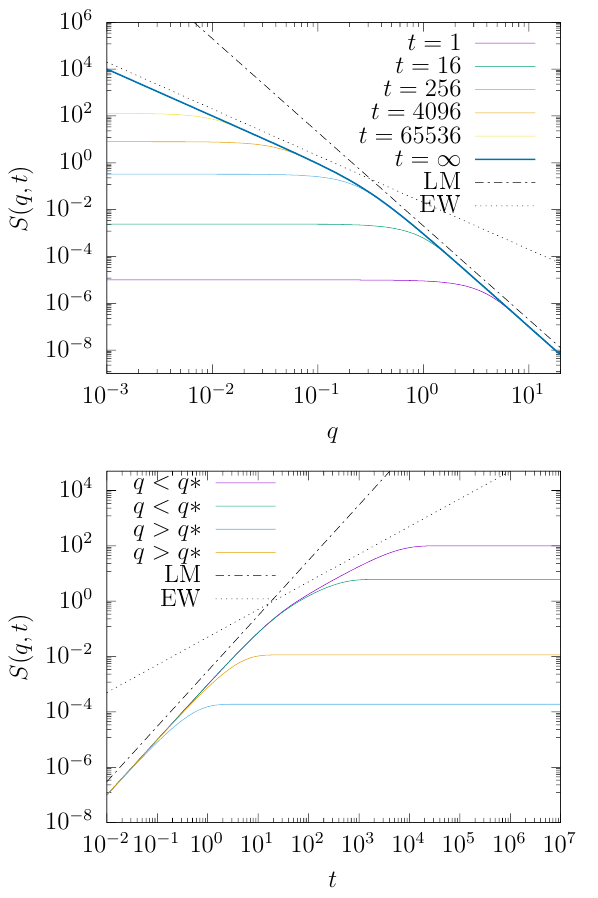}
\caption{
Analytical predictions for the Larkin (LM) to Edwards-Wilkinson (EW) crossover during the relaxation of a flat initial interface, for $\tau=10$.
(a) $S(q,t)$ vs $q$ at different times. The dotted line shows EW scaling $\sim q^{-2}$ and the dashed-dotted the LM scaling $S(q,t) \sim q^{-(1+2\zeta_{\rm LM})}=q^{-4}$. (b) $S(q,t)$ vs $t$ for $q>q^*$ and $q<q^*$, with $q^*$ the crossover wavevector between LM to EW. The plateau at low $q$ shows non-equilibrated modes with memory of the initial flat condition.
The dotted line shows EW scaling $S(q,t) \sim t^{1/z_{\rm EW}}=t^{1/2}$ and the dashed-dotted the LM scaling $S(q,t) \sim t^{(1+2\zeta_{\rm LM})/z_{\rm LM}}=t^{2}$. The onset of the plateau in each case marks the equilibration time and amplitude of the corresponding mode $q$, below and above the crossover $q^*$.
}
\label{fig:analyticalLMtoEW}
\end{figure}

Before addressing the nonlinear problem, we analyze the exactly solvable linear limit $c_{2n}=0$. This case already exhibits the crossover between Larkin and Edwards--Wilkinson scaling induced by the finite persistence time $\tau$, and provides the natural reference theory for the nonlinear analysis. For simplicity, we restrict ourselves to $d=1$.

\subsection{Structure Factor}
We rewrite Eqs.~\eqref{eq:eqmotion} and \eqref{eq:colorednoise} as an enlarged Markovian system for the fields $h$ and $\eta$:
\begin{align}
\partial_t h(x,t) &= c_2 \partial_x^2 h(x,t) + \eta(x,t), \nonumber \\
\tau \partial_t \eta(x,t) &= -\eta(x,t) +\xi(x,t), \nonumber \\
\langle \xi(x,t)\,\xi(x',t') \rangle &= 2T\delta(x-x')\,\delta(t-t'),
\label{eq:linearmodel}
\end{align}
which becomes diagonal in Fourier space. Using the convention
\[
h(x)=\frac{1}{\sqrt{L}}\sum_q h_q\,e^{iqx},
\qquad
\eta(x)=\frac{1}{\sqrt{L}}\sum_q \eta_q\,e^{iqx},
\]
we obtain the stationary structure factor
\begin{equation}
\langle |h_q|^2 \rangle = \frac{T}{c_2 q^2\left(1+\tau c_2 q^2\right)} \equiv S(q).
\label{eq:sofqLMtoEW}
\end{equation}

This expression exhibits a crossover from a Larkin model regime (LM), $S(q)\sim \frac{T}{c_2 \tau} q^{-4}$, to an Edwards--Wilkinson model regime (EW), $S(q)\sim \frac{T}{c_2} q^{-2}$, at the characteristic wavevector $q_{\rm LM/EW} = (c_2 \tau)^{-1/2}$. Therefore, the crossover length scale is
\begin{equation}
l_{\rm LM/EW}=2\pi\sqrt{c_2\tau}.
\end{equation}
This is the characteristic length scale introduced by the finite persistence time, used to normalize lengths in Fig. \eqref{fig:crossoverdiagram} as indicated in Eq. \eqref{eq:adimensionallength}.

In the infrared limit ($l \gg l_{\rm LM/EW}$), the structure factor is insensitive to the non-Markovian character of the finite-$\tau$ dynamics, and the EW scaling is identical to that predicted for the equilibrium $\tau \to 0$ case. Within the linear theory, the LM/EW crossover is the only crossover expected for any $\tau > 0$.

Assuming a flat initial condition $h(x,0)=0$, the non-stationary structure factor, for e.g. a flat initial condition $h(x,0)=0$, can also be obtained analytically:

\begin{eqnarray}
S(q,t) &=
\frac{T}{c_2 q^2\left(1+c_2 q^2\tau\right)}
\left(1-e^{-2c_2 q^2 t}\right) \nonumber \\
&-
\frac{2T\tau}
{1-c_2^2 q^4\tau^2}
\left(
e^{-2c_2 q^2 t}
-
e^{-(c_2 q^2+1/\tau)t}
\right).
\label{eq:sofqLMtoEWvst}
\end{eqnarray}

For $t \gg \max\{\tau, (c_2 q^2)^{-1}\}$, this converges to $S(q)$ in Eq.~\eqref{eq:sofqLMtoEW}. Two dynamical scaling regimes are identified. For short times $t\ll\tau$, $S(q,t)= q^{-(1+2\zeta_{\rm LM})}\, F(q\, t^{1/z_{\rm LM}})$, while for $t\gg\tau$ and $q\,l_{\rm LM/EW}<1$, $S(q,t)= q^{-(1+2\zeta_{\rm EW})}\, G(q\, t^{1/z_{\rm EW}})$.

As shown in Fig.~\ref{fig:analyticalLMtoEW}, the modes relax towards their stationary amplitudes in a wavevector-dependent manner. Modes with $q<q_{\rm LM/EW}$ grow as $S(q,t)\sim t^2$ before saturation, while modes with $q>q_{\rm LM/EW}$ display a crossover from $t^2$ to $t^{1/2}$ growth. Consequently, the global roughness evolves from $w^2(t)\sim t^{3/2}$ to $w^2(t)\sim t^{1/2}$ on timescales larger than $\tau$.

\subsection{Effective temperatures}

The two-time correlation function
\(
C(q,t,t')=\langle h_q(t)h_{-q}(t')\rangle
\)
and the linear response function
\begin{equation}
R(q,t,t')
=
\Theta(t-t')e^{-c_2 q^2(t-t')}
\end{equation}
allow one to define a generalized fluctuation--dissipation ratio. Away from equilibrium, this motivates the definition of a time-dependent effective temperature ~\cite{Cugliandolo_2011}
\begin{equation}
T_{\rm eff}(q,t,t')=
\frac{\partial_{t'}C(q,t,t')}
{R(q,t,t')}.
\end{equation}

In the stationary regime, where both correlation and response depend only on the time difference, it is more convenient to work in frequency space. The effective temperature is then defined through
\begin{equation}
T_{\rm eff}(q,\omega)
=
\frac{\omega C(q,\omega)}
{2(-\operatorname{Im}R(q,\omega))}.
\end{equation}

For the linear theory with exponentially correlated noise,
\begin{equation}
R(q,\omega)
=
\frac{1}{-i\omega+c_2 q^2},
\end{equation}
while the correlation spectrum is
\begin{equation}
C(q,\omega)
=
\frac{2T}{1+\omega^2\tau^2}
\frac{1}{(c_2 q^2)^2+\omega^2},
\end{equation}
then 
\begin{equation}
T_{\rm eff}(q,\omega)
=
\frac{T}{1+\omega^2\tau^2}.
\end{equation}
The effective temperature is therefore independent of the wavevector and depends only on frequency. In the low-frequency regime,
\begin{equation}
\omega\tau\ll1,
\qquad
T_{\rm eff}\simeq T,
\end{equation}
so the system behaves effectively as if it were in thermal equilibrium. In contrast, for
\begin{equation}
\omega\tau\gg1,
\end{equation}
the effective temperature decays as
\begin{equation}
T_{\rm eff}\sim \frac{T}{\omega^2\tau^2},
\end{equation}
reflecting the suppression of fast fluctuations by the finite noise correlation time. 
These high-frequency modes correspond to the short-length-scale Larkin sector of the interface, whose relaxation time $(c_2 q^2)^{-1}$ is shorter than the persistence time $\tau$. They therefore experience the active noise as an effectively frozen random force.

\subsection{Mapping to a stiff interface}
An additional consequence of the finite correlation time is that, for Gaussian noise, the steady-state statistics are completely determined by the structure factor.
Specifically, the probability functional for the interface configurations $h$ is given by the multivariate Gaussian distribution $P_s[h] \propto \exp [ -\frac{1}{2}\sum_q |h_q|^2 / S(q) ]$. By substituting Eq.~\eqref{eq:sofqLMtoEW} into this expression and transforming back to real space, we find that the stationary measure is:
\begin{eqnarray}
P_s[h]
&\propto& 
\exp\Big\{
\frac{-1}{2T} \int dx 
\left[ 
c_2 (\partial_x h)^2 + \tau c_2^2 (\partial_x^2 h)^2
\right]
\Big\}.
\end{eqnarray}
Interestingly, in the steady state, the finite persistence $\tau$ introduces an effective bending rigidity term $\propto (\partial_x^2 h)^2$ into the Hamiltonian. This establishes, in the steady-state, a formal mapping between the active linear interface and the equilibrium statistics of a stiff interface \cite{ChaikinLubensky1995}.

\section{Nonlinear roughening and crossovers}

We now consider the full nonlinear model with $c_{2n}>0$. In contrast to the linear theory, no exact analytical solution is available. To characterize the steady-state roughening, we combine scaling arguments with a self-consistent Hartree approximation. This approach allows us to derive the crossover diagram introduced in Fig.~\ref{fig:crossoverdiagram}, determine the associated crossover length scales, and obtain analytical predictions for the structure factor across the different scaling regimes. The resulting theory is tested against numerical simulations in the next section.

\subsection{Multiscale crossover picture: self-affine regimes}
\label{sec:multiscale}

The crossover structure of Eq.~\eqref{eq:eqmotion} is controlled by the competition between the persistence time $\tau$ and the relaxation times of interface modes. Modes with relaxation times shorter than $\tau$ experience the active forcing as effectively quenched, whereas modes with relaxation times longer than $\tau$ probe the time-averaged noise and recover Edwards--Wilkinson behavior.

Among the fast modes, a second crossover arises from the competition between harmonic and nonlinear elasticity. At short length scales harmonic elasticity dominates and the interface follows Larkin scaling. Since $\zeta_{\rm LM}>1$ in $d=1$, power counting shows that the anharmonic elastic term is relevant and destabilizes the LM fixed point. Beyond a characteristic length $l_{\rm anh}$, the dynamics therefore crosses over to the anharmonic Larkin model (ALM)~\cite{purrello2019roughening}.

At large length scales, where mode relaxation times exceed $\tau$, the active noise is effectively white. Power counting around the EW fixed point shows that the anharmonic elastic term is irrelevant for any $n>1$, so the asymptotic large-scale behavior remains Edwards--Wilkinson-like, although with renormalized amplitudes.

We will  argue that the three regimes, LM, ALM and EW, are relevant for describing the steady-state solutions of Eq. \eqref{eq:eqmotion}. In the following we describe the minimal equations of motion that generate each one and their critical exponents. 

\begin{itemize}
\item \textbf{Larkin regime (LM).}  
This regime is predicted to occur for 
$l \ll l_{\rm anh}$.
At very short times, linear elasticity dominates and the noise is effectively quenched. 
The effective equation can be approximated as
\begin{equation}
 \partial_t  h(x,t) \approx c_2 \partial_x^2 h(x,t) + \eta(x,0),
\label{eq:eqmotionLM}
\end{equation}
and scale invariance under $x\to bx$, $t\to b^z t$, $h\to b^\zeta h$ 
yield the global exponents
\[
\zeta_{\rm LM} = \frac{4 - d}{2}, \qquad z_{\rm LM} = 2.
\]
such that $\langle h^2 \rangle \sim L^{2\zeta_{\rm LM}}$, with $L$ the linear size, after a typical equilibration time of order $L^{z_{\rm LM}}$. While $z_{\rm LM}$ is independent of $d$, 
the roughness exponent $\zeta_{\rm LM}$ is valid only for $d<4$, and zero otherwise.
The LM fixed point preserves statistical tilt symmetry and is unstable in $d=1$ because $\zeta_{\rm LM}>1$, making anharmonic elasticity relevant.

\item \textbf{Anharmonic Larkin regime (ALM).}  
Since $\zeta_{\rm LM}(d=1)>1$, for $\tau_{\rm anh} < t < \tau$ the one dimensional interface is expected to cross over from LM to a regime dominated by anharmonic elasticity.
The effective dynamics in this regime can be approximated as
\begin{equation}
 \partial_t  h(x,t) = c_{2n} \partial_x \!\left[(\partial_x h(x,t))^{2n-1}\right] + \eta(x,0),
\label{eq:eqmotionALM}
\end{equation}
and rescaling predicts the global exponents
\begin{equation}
\zeta_{\rm ALM} = \frac{4n-d}{4n-2}, \qquad
z_{\rm ALM} = \zeta_{\rm ALM} + \frac{d}{2},    
\end{equation}
such that $\langle h^2 \rangle \sim L^{2\zeta_{\rm ALM}}$, with $L$ the linear size, after a typical equilibration time of order $L^{z_{\rm ALM}}$.
The $n=1$ case reduces to the previous LM case, and  it is excluded from this class. 
For $n>1$ and $d=1$ the scaling is however anomalous, and the spectral exponent is larger than the global exponent, $\zeta_{\rm s} > \zeta_{\rm ALM} \geq 1$ for $d=1$, such that $S(q) \sim q^{-(1+2\zeta_{\rm s})}$. 
Interestingly, the roughening in this case is characterized by the growth of facets
\cite{purrello2019roughening}.
This model preserves the $h \to -h$ symmetry.
Unlike the LM and EW fixed points however, the $n>1$ ALM  dynamics breaks statistical tilt symmetry.

\item \textbf{Edwards--Wilkinson regime (EW).}  
For $\tau < \tau_{\rm anh} < t$, the noise decorrelates before anharmonic effects become important.  
The dynamics is then effectively described by
\begin{equation}
 \partial_t  h(x,t) = c_2 \partial_x^2 h(x,t) + \xi(x,t),
\label{eq:eqmotionEW}
\end{equation}
where $\xi(x,t) \equiv \lim_{\tau \to 0} \eta(x,t)$ is uncorrelated white noise.  
Scaling arguments suggest
\[
\zeta_{\rm EW} = \frac{2 - d}{2}, \qquad z_{\rm EW} = 2.
\]
While $z_{\rm EW}$ is independent of $d$, 
the roughness exponent $\zeta_{\rm EW}$ is valid only for $d<2$, and zero otherwise.
This model respect the statistical tilt symmetry, $h \to h-s x$ with $s$ some slope and the inversion simmetry $h \to -h$.
\end{itemize}

The EW regime can be thus reached either directly from the LM regime when $\tau<\tau_{\rm anh}$ or after an intermediate ALM regime when $\tau>\tau_{\rm anh}$, where $\tau_{\rm anh}$ is the characteristic $\tau$ at which the crossover length to the EW regime coincides with the anharmonic (LM to ALM) crossover length $l_{\rm anh}$  of the quenched ALM model \cite{purrello2019roughening},

\begin{equation}
l_{\rm anh}={\left(\frac{c_2^2 \tau}{T}\right)^{\frac{1}{2-d}}} 
             {\left(\frac{n c_2}{c_{2n}}\right)}^{\frac{1}{(2-d)(n-1)}}
             \equiv l_1 \left(\frac{\tau}{\tau_1}\right)^{\frac{1}{2-d}},
\label{eq:lanhvictor}
\end{equation}

with $l_1$ and $\tau_1$ $\tau$-independent characteristic length and time scales, respectively, 

\begin{equation}
    \tau_1 = \left( \frac{c_2^d}{T} \right)^{\frac{2}{d}} \left( \frac{n c_2}{c_{2n}} \right)^{\frac{2}{d(n-1)}},
\quad
    l_1 = \sqrt{c_2 \tau_1}. 
\end{equation}
At the characteristic time scale $\tau_1$ the active persistence  ``activates'' the nonlinearities. We then deduce that
\begin{equation}
\tau_{\rm anh} \sim \tau_1 (l_{\rm anh}/l_1)^{z_{\rm LM}}    
\label{eq:tauanhvictor}
\end{equation}

\begin{table}[h]
\caption{\label{tab:exponents} 
Scaling exponents for the three roughening regimes discussed in the text.
Roughness exponents are valid for $d < d_c$. In the ALM, scaling is anomalous and the spectral exponent is $\zeta_{\rm s} > \zeta_{\text{ALM}}$. \cite{purrello2019roughening}.}
\begin{tabular}{|l|c|c|c|}
\hline
Model & $\zeta$ & $z$ & $d_c$ \\
\hline
LM & $\displaystyle \zeta_{\text{LM}} = \frac{4 - d}{2}$ & $\displaystyle z_{\text{LM}} = 2$ & 4 \\[1.5ex]
\hline
ALM & $\displaystyle \zeta_{\text{ALM}} = \frac{4n - d}{4n - 2}$ & $\displaystyle z_{\text{ALM}} = \zeta_{\text{ALM}} + \frac{d}{2}$ & $4n$ \\[1.5ex]
\hline
EW & $\displaystyle \zeta_{\text{EW}} = \frac{2 - d}{2}$ & $\displaystyle z_{\text{EW}} = 2$ & 2 \\
\hline
\end{tabular}
\label{tab:exponents3regimes}
\end{table}

\subsection*{Crossover lengths and diagram}

A crucial point is to realize that $l_{\rm anh}$ (Eq. \eqref{eq:lanhvictor}) and $\tau_{\rm anh}$ (Eq. \eqref{eq:tauanhvictor}) are $\tau$ dependent because $\eta(x,0)$ has an effective $\tau$ dependent amplitude $\sqrt{T/\tau}$. Nevertheless, the crossover lengths separating the different regimes can be predicted as
\begin{align}
l_{\rm LM/EW}  &= l_0 \left(\frac{\tau}{\tau_0}\right)^{1/z_{\rm LM}}, & & \tau/\tau_{\rm anh} < 1, 
\label{eq:lmtoew}
\\
l_{\rm LM/ALM} &= l_{\rm anh} = l_0 \left(\frac{\tau_{\rm anh}}{\tau_0}\right)^{1/z_{\rm LM}}, & & \tau/\tau_{\rm anh} > 1,
\label{eq:lmtoalm}
\\
l_{\rm ALM/EW} &= l_{\rm anh} \left(\frac{\tau}{\tau_{\rm anh}}\right)^{1/z_{\rm ALM}}, & & \tau/\tau_{\rm anh} > 1,
\label{eq:almtoew}
\end{align}
where $l_0$ and $\tau_0$ are microscopic length and time scales which are independent of $\tau$, $T$ and elastic constants $c_2$ and $c_{\rm 2n}$.  
All three crossover lines (Eqs.\eqref{eq:lmtoew}, \eqref{eq:lmtoalm}, \eqref{eq:almtoew}) intersect at
$(\tau_{\rm anh}, l_{\rm anh})$.

Note that in $d=1$, $l_{\rm anh} \sim l_1(\tau/\tau_1)$ and $\tau_{\rm anh} \sim \tau_1 (l_{\rm anh}/l_1)^{z_{\rm LM}} \sim \tau^{2}$. 
Therefore, for $d=1$,
\begin{align}
l_{\rm LM/EW} &\sim \tau^{1/2},& & \tau \gg \tau_1\\ 
l_{\rm LM/ALM} &\sim \tau^{\frac{n}{2(n-1)}} ,& & \tau \ll \tau_1\\
l_{\rm ALM/EW} &\sim \tau^{1-1/z_{\rm ALM}} = \tau^{\frac{n}{3n-1}}
,& & \tau \ll \tau_1
\label{eq:crossoverlinesfrommultiscaling}
\end{align}
We note that 
since $l_{\rm ALM/EW} \geq l_{\rm LM/ALM}$ should be necessarily satisfied, 
the ALM regime disappears for large $\tau$ because in that case the short length scale LM regime crossovers directly to the EW regime. 
These scaling arguments reproduce the topology of the crossover diagram shown in Fig.~\ref{fig:crossoverdiagram} and provide the scaling forms of the crossover lines for fixed $T$, $c_2$, and $c_{2n}$.

\subsection{Self-consistent Hartree approximation}
\label{sec:hartree}

Although uncontrolled, the Hartree approximation provides a simple nonperturbative estimate of how nonlinear elasticity renormalizes the effective surface tension and therefore the crossover scale to the EW regime.

To estimate the crossover to the asymptotic EW regime, we replace the nonlinear elasticity by a self-consistent effective surface tension $\nu_{\rm eff}$. The resulting Hartree approximation maps Eq.~\eqref{eq:eqmotion} onto the linear problem
\begin{equation}
 \partial_t h(x,t)
 =
 \nu_{\rm eff}\partial_x^2 h(x,t)
 +
 \eta(x,t),
\label{eq:coarsegrainedeqmotion}
\end{equation}
where $\nu_{\rm eff}$ is determined self-consistently.

Assuming Gaussian slope fluctuations, Wick's theorem yields
:
\begin{equation}
    \nu_{\rm eff} = c_2 + c_{2n} (2n-1)!! (\sigma^2)^{n-1},
\end{equation}
where $\sigma^2 = \langle (\partial_x h)^2 \rangle$. Using the result of 
Eq. \eqref{eq:sofqLMtoEW}, replacing $c_2 \to \nu_{\rm eff}$ we get
\begin{equation}
\sigma^2 = \frac{T}{\nu_{\text{eff}}} \int_{-\infty}^{\infty} \frac{dk}{2\pi} \frac{1}{1 + \nu_{\text{eff}} \tau k^2} =
\frac{T}{2 \sqrt{\tau} (\nu_{\text{eff}})^{3/2}}.
\label{eq:integralhartree1}
\end{equation}
The finite persistence time suppresses modes with
$k \gtrsim (\nu_{\rm eff}\tau)^{-1/2}$,
so the integral is naturally controlled by the scale
$
\Lambda \sim (\nu_{\rm eff}\tau)^{-1/2}.
$
This defines the crossover length
\begin{equation}
\ell = \sqrt{\nu_{\rm eff}\tau}.    
\end{equation}
Then, the self consistent equation is
\begin{equation}
\nu_{\text{eff}} - c_{2n} \frac{(2n-1)!!}{(2\sqrt{\tau})^{n-1}} T^{n-1} (\nu_{\text{eff}})^{-\frac{3}{2}(n-1)} = c_2,
\label{eq:self_cons_naive}
\end{equation}
and can be used to estimate $\nu_{\text{eff}}$ and $\ell$.

Analyzing the self-consistent equation in the limits of small and large persistence times yields the scaling behavior summarized in Table~\ref{tab:hartreeregimes}. In the regimes $\tau\to0$ or $c_2\ll \nu_{\rm eff}-c_2$, the characteristic length $\ell=\sqrt{\nu_{\rm eff}\tau}$ scales as the crossover length $l_{\rm ALM/EW}$ obtained from the multiscale arguments of Sec.~\ref{sec:multiscale}. Conversely, for $\tau\to\infty$, one recovers $\nu_{\rm eff}\to c_2$ and $\ell\sim l_{\rm LM/EW}$. 
Remarkably, the crossover exponents obtained from the Hartree closure coincide with those predicted by the scaling analysis of Sec.~\ref{sec:multiscale}. 

The surprisingly good performance of the Hartree approximation is reminiscent of previous observations in disordered elastic interfaces~\cite{ledoussal2003,Rosso2003,Moulinet2004}, where interface fluctuations were found to be accurately described by generalized Gaussian theories despite the presence of strong nonlinear correlations.

\begin{table}[h!]
\centering
\begin{tabular}{|l|c|c|c|}
\hline
\textbf{Condition} & \textbf{Effective Tension } $\nu_{\text{eff}}$ & \textbf{Length Scale } $\ell$ \\ \hline
$\tau \to 0$ & $\nu_{\text{eff}} \sim \tau^{-\frac{n-1}{3n-1}}$ & $\ell \equiv l_{\rm ALM/EW} \sim \tau^{\frac{n}{3n-1}}$ \\ \hline
$c_2 \ll \nu_{\rm eff}-c_2$ & $\nu_{\text{eff}} \sim \tau^{-\frac{n-1}{3n-1}}$ & $\ell \equiv l_{\rm ALM/EW} \sim \tau^{\frac{n}{3n-1}}$ \\ \hline
$\tau \to \infty$ & $\nu_{\text{eff}} \to c_2$ & $\ell \equiv l_{\rm LM/EW} \sim \tau^{1/2}$ \\ \hline
\end{tabular}
\caption{
Scaling of the Hartree effective tension and crossover length for $n\ge2$. The limits $\tau\to0$ and $c_2\ll\nu_{\rm eff}-c_2$ correspond to the nonlinear regime, while $\tau\to\infty$ recovers the linear LM--EW crossover.
}
\label{tab:hartreeregimes}
\end{table}

The Hartree approximation confirms the scaling predictions for the crossover to the EW regime, but does not resolve the LM--ALM crossover. We now combine the Hartree approach with the multiscale arguments of Sec.~\ref{sec:multiscale} to derive the complete crossover diagram and its dependence on $T$, $\tau$, $c_2$, and $c_{2n}$.

\subsection{Structure Factor general scaling}
\label{sec:generalscalings}

We now combine the multiscale arguments of Sec.~\ref{sec:multiscale} with the Hartree analysis of Sec.~\ref{sec:hartree} to derive a general scaling description of the steady-state structure factor. The central result is that, for a given nonlinearity order $n$, the crossover diagram is controlled by a single dimensionless parameter $\Gamma_{2n}$, which combines the persistence time, noise amplitude, and elastic constants. This immediately yields the scaling form of the crossover lengths and the structure factor.

Although the remainder of the paper focuses on a single nonlinear coupling $c_{2n}$, it is convenient to formulate the scaling analysis for the more general elastic energy containing arbitrary even powers of the local slope:
\begin{equation}
  \partial_t h(x,t) = \partial_x\!\left(\sum_{n=1}^{\infty}
  c_{2n}\,\bigl(\partial_x h(x,t)\bigr)^{2n-1}\right) + \eta(x,t),
  \label{eq:model}
\end{equation}
where $\eta$ is given 
by Eq. \eqref{eq:colorednoise}. This equation reduces to \eqref{eq:eqmotion} when we keep the terms $n=1$ and $n=m>1$ for a given integer $m$. 

We first introduce rescaled variables:
\begin{equation}
  y = \frac{x}{\gamma\sqrt{\tau}}, \qquad
  s = \frac{t}{\tau}, \qquad
  u(y,s) = \frac{h(x,t)}{\sqrt{\tau}},
  \qquad \tilde{\eta} = \sqrt{\tau}\,\eta
  \label{eq:change1}
\end{equation}
where $\gamma$ is a free parameter. 
Such change of variables leads to 
\begin{equation}
  \partial_s u = \partial_y\!\left(\sum_{n=1}^{\infty}
  \frac{c_{2n}}{\gamma^{2n}}\,(\partial_y u)^{2n-1}\right)
  + \tilde{\eta}(y,s).
  \label{eq:utilde}
\end{equation}
The transformed noise correlator is
 \begin{align} 
 \langle\tilde{\eta}(y,s)\,\tilde{\eta}(y',s')\rangle
  &= \frac{T}{\gamma\sqrt{\tau}}\,\delta(y-y')\,e^{-|s-s'|}
  =\nonumber \\ 
  &\tilde{T}\,\delta(y-y')\,e^{-|s-s'|},
  \label{eq:noise_utilde}
  \end{align}
where the rescaled temperature is

\begin{equation}
 \tilde{T} = \frac{T}{\gamma\sqrt{\tau}}.
  \label{eq:Teff}
\end{equation}
Equation~\eqref{eq:utilde}is identical in form to the original equation~\eqref{eq:model} with $\tau=1$ and temperature $\tilde{T}$.

We now define a rescaled displacement field
\begin{equation}
  w(y,s) = \frac{u(y,s)}{\sqrt{T_{\mathrm{eff}}}}
  = h(x,t)\sqrt{\frac{\gamma}{T\sqrt{\tau}}},
  \label{eq:change2}
\end{equation}
Then, substituting into Eq.~\eqref{eq:utilde} and dividing by
$\sqrt{T_{\mathrm{eff}}}$, the equation for $w$ is
\begin{equation}
  \partial_s w = \partial_y\!\left(\sum_{n=1}^{\infty}
  \lambda_{2n}\,(\partial_y w)^{2n-1}\right) + \hat{\eta}(y,s),
  \label{eq:w}
\end{equation}
with effective couplings renormalized by the effective temperature:
\begin{equation}
  \lambda_{2n} = \frac{c_{2n}\,T_{\mathrm{eff}}^{\,n-1}}{\gamma^{2n}}
  = \frac{c_{2n}}{\gamma^{3n-1}}\left(\frac{T}{\sqrt{\tau}}\right)^{\!n-1},
  \label{eq:lambda}
\end{equation}
\color{black}
and a noise
\begin{equation}
  \langle\hat{\eta}(y,s)\,\hat{\eta}(y',s')\rangle
  = \delta(y-y')\,e^{-|s-s'|},
  \label{eq:noise_w}
\end{equation}
which is independent of $T$, $\tau$, and $\gamma$.

Note that:
\begin{itemize}
  \item $\lambda_2 = c_2/\gamma^2$: the linear elasticity does not renormalize
with $\tilde{T}$
  \item $\lambda_{2n} = (c_{2n}/\gamma^{2n})\,\tilde{T}^{n-1}$: each nonlinear coupling scales with a higher power of $\tilde{T}$.
\end{itemize}

The advantage of the transformation of the equation of motion \eqref{eq:eqmotion} and \eqref{eq:colorednoise} into Eqs. \eqref{eq:w}, \eqref{eq:lambda}, \eqref{eq:noise_w} is that all dependence on $T$, $\tau$ and $n$ is now contained in a single parameter and that the free parameter $\gamma$ can be chosen to simplify the effective couplings.
In particular, the steady-state structure factor 
can be written as 
\begin{align}
  S(q\,|\,T,\tau)
  &= T\tau\cdot
     \int dy\,e^{iq\gamma\sqrt{\tau}\,y}\,\langle w(y)\,w(0)\rangle
     \nonumber\\
  &= T\tau\cdot\tilde{S}\!\left(q\gamma\sqrt{\tau}\,\Big|\,
     \left\{\lambda_{2n}\right\}\right)
  = T\tau\cdot\tilde{S}\!\left(k\,\Big|\,
     \left\{\lambda_{2n}\right\}\right),   
     \label{eq:Sdekylambdas}
\end{align}
where we defined $k=q\gamma \sqrt{\tau}$. In terms of the original variables we obtain the general result:
\begin{equation}
  S(q\,|\,T,\tau) = T\tau\cdot\tilde{S}\!\left(q\gamma\sqrt{\tau}\,
  \Big|\,\left\{\frac{c_{2n}}{\gamma^{2n}}
  \left(\frac{T}{\gamma\sqrt{\tau}}\right)^{n-1}\right\}\right)
  \label{eq:S_general}
\end{equation}
where $\gamma$ is still free and we will use it to analyze three different cases of interest. 

\subsubsection{Linear case: only \texorpdfstring{$c_2\neq 0$}{c2}}

Choosing $\gamma^2 = c_2$ so that $\lambda_2=1$ in Eq.\eqref{eq:Sdekylambdas}, Eq. \eqref{eq:S_general} gives:
\begin{equation}
        S(q\,|\,T,\tau) = T\tau\cdot f(q\sqrt{c_2\tau}),
  \label{eq:S_linear_0}
\end{equation}
which is indeed the form given by the analytical exact solution Eq. \eqref{eq:sofqLMtoEW}.
Since the argument of $f$ is non-dimensional, the characteristic wavevector  
$q_c \sim 1/\sqrt{c_2\tau}$ can be identified with  the crossover length $l_{\rm LM/EW} \sim \tau^{1/2}$, between the LM to EW regime.

\subsubsection{Purely nonlinear case:  \texorpdfstring{$c_{2n}\neq 0$}{c2n} (single \texorpdfstring{$n>1$}{n>1})}

Choosing $\gamma$ so that $\lambda_{2n}=1$ for a single $n>1$,
\begin{equation}
  \gamma = \left(\frac{c_{2n}\,T^{n-1}}{\tau^{(n-1)/2}}\right)^{1/(3n-1)},
\end{equation}
the structure factor depends only on the adimensional quantity
\begin{equation}
  k = q\left(c_{2n}\,T^{n-1}\,\tau^n\right)^{1/(3n-1)},
\end{equation}
We can then identify a characteristic 
scale
$q_c = \left(c_{2n}\,T^{n-1}\,\tau^n\right)^{-1/(3n-1)}$, which can be associated to the expected ALM to EW crossover 
$l_{\rm ALM/EW} \equiv q_c^{-1} \sim \tau^{n/(3n-1)}$ (see Sec. \ref{sec:multiscale}).
In particular, for the lowest order pure nonlinear elasticity ($n=2$) we get,
\begin{equation}
        S(q\,|\,T,\tau) = T\tau\cdot f (q (c_4 T)^{1/5}\tau^{2/5}),
  \label{eq:S_c4}
\end{equation}
implying a crossover wavevector scaling as $q \sim (c_4 T)^{-1/5} \tau^{-2/5}$, which can be identified with the crossover length $l_{\rm ALM/EW} \sim \tau^{2/5}$ between the ALM to EW regimes.

\subsubsection{Mixed case: \texorpdfstring{$c_2$}{c2} and \texorpdfstring{$c_{2n}$}{c4} nonzero}

Choosing $\gamma^2=c_2$ so that $\lambda_2=1$ and
$\lambda_{2n}=\Gamma_{2n}\equiv c_{2n}T/(c_2^{(3n-1)//2}\sqrt{\tau})$ we get,
\begin{equation}
  S(q\,|\,T,\tau) = T\tau\cdot f\!\left(q\sqrt{c_2\tau},\,\Gamma_{2n}\right),
  \label{eq:S_mixed}
\end{equation}
where 
\begin{equation*}
\color{black}
\boxed{\Gamma_{2n} = \frac{c_{2n}\,T}{c_2^{(3n-1)/2}\sqrt{\tau}}}
\label{eq:Gamma2n}
\color{black}
\end{equation*}
controls the crossover diagram in Fig.~\ref{fig:crossoverdiagram} (cf. Eq. \eqref{eq:gamma_coupling_definition}).

In this case the shape of $f$ is parametrized by the dimensionless parameter $\Gamma_{2n}$. Using the arguments of Sec. \ref{sec:multiscale} 
we can anticipate two limiting cases for $f$:
\begin{itemize}
  \item $\Gamma_{2n}\ll 1$: two regimes EW-LM, crossover at $k\sim 1$
  \item $\Gamma_{2n}\gg 1$: three regimes EW-ALM-LM, with two crossovers
        $k_1(\Gamma_{2n})$ and $k_2(\Gamma_{2n})$, both depending on $\Gamma_{2n}$
\end{itemize}
In the transformed variables that lead to Eq. \eqref{eq:w} from Eq. \eqref{eq:lanhvictor}, using that we have $\tilde c_2=1$, $\tilde c_4 = \Gamma$,  $\tilde \tau = 1$, $
\tilde T=1$, we can then deduce the transformed anharmonic crossover length, 
\color{black}$\tilde l_{\rm anh} \sim (\Gamma_{2n}/n)^{-1/(n-1)}$,
\color{black}

and then the scaling of all possible crossover lengths as a function of the relative coupling parameter $\Gamma_{2n}$ are 
\begin{eqnarray}
    \tilde l_{\rm LM/EW} &\sim \text{const},\quad &\Gamma_{2n} \ll 1, \\
    \tilde l_{\rm ALM/EW} &\sim (\Gamma_{2n}/n)^{1/(3n-1)}, \quad &\Gamma_{2n} \gg 1, \\ 
    \tilde l_{\rm LM/ALM} &\equiv \tilde l_{\rm anh} \sim 
    \color{black}(\Gamma_{2n}/n)^{-1/(n-1)} 
    \color{black}
    \quad &\Gamma_{2n} \gg 1.
\end{eqnarray}
From these scalings we deduce that the ALM regime can only exist above a finite threshold $\Gamma_{2n} \sim n$ so that the nonlinear elasticity can modify the roughness of modes whose relaxation times are faster than $\tilde \tau$.

The condition
$l_{\rm ALM/EW}>l_{\rm LM/ALM}$
implies the existence of a finite threshold
$\Gamma_{2n}\sim n$.
Below this threshold the LM regime crosses directly into the EW regime and no intermediate ALM scaling window exists. These results directly support the crossover diagram of Fig.~ \ref{fig:crossoverdiagram}.

\subsubsection*{Structure Factor}
The crossover diagram determines not only the crossover lengths but also the full structure factor. Since the large-scale regime is always Edwards--Wilkinson, it is natural to factor out the EW contribution and express the remaining scale dependence through a universal crossover function.

Let us start writing the large scale EW behavior as
\begin{equation}
        S\sim \frac{T}{\nu_{\mathrm{eff}}q^2}=\frac{T\tau}{\tilde\nu_{\mathrm{eff}}k^2},\;\;\;\;\;\;\text{for}\;\;\;\;\; q\to 0
\end{equation}
where $k=q\sqrt{c_2\tau}$ and we defined a nondimensional effective elastic constant $\tilde\nu_{\mathrm{eff}}=\nu_{\mathrm{eff}}/c_2$ (which depends only on $\Gamma_{2n}$).
We then express the structure factor in the form

\begin{equation}
S(q\,|\,T,\tau) = \frac{T\tau}{\tilde\nu_{\mathrm{eff}}k^2} g\!\left(k,\,\Gamma_{2n}\right)
\end{equation}

According to the discussion above, we expect two different behaviors for the universal crossover function $g(k,\Gamma_{2n})$

\begin{itemize}
        \item
                $\Gamma_{2n}<n$: LM-EW  crossover

                \begin{equation}
                        g_<(k,\Gamma_{2n})=
                        \left\{
                        \begin{array}{lcr}
                                1 &\text{if} & k< 1\\
                                k^{-2} &\text{if}& k > 1
                        \end{array}
                        \right.
                        \label{eq:g<}
                \end{equation}
                Thus, $g_<$ does not depend on $\Gamma_{2n}$ 
                
        \item
                $\Gamma_{2n}>n$ LM-ALM and ALM-EW crossovers:
                 \begin{equation}
                        g_>(k,\Gamma_{2n})=
                        \left\{
                        \begin{array}{lcr}
                                1 &\text{if} & k< k_1\\
                                &&\\
                                k_1^{(2\zeta_s-1)}\cdot k^ {-(2\zeta_s-1)}&\text{if}& k_1< k < k_2\\
                                &&\\
 k_1^{(2\zeta_s-1)} k_2^{-(2\zeta_s-3)}\cdot k^ {-2} & \text{if}&     k_2 < k
                        \end{array}
                        \right.
                        \label{eq:g>}
                \end{equation}

\end{itemize}
Where

\begin{equation}
\color{black}
    k_1= (\Gamma_{2n}/n)^{-1/(3n-1)},\qquad k_2=(\Gamma_{2n}/n)^{1/(n-1)}
\color{black}
\label{eq:k1andk2}
\end{equation}
In both cases we assumed a sharp crossover between all regimes at the characteristic crossover lengths.\\

\subparagraph{\underline{$n=2$}}

The scaling theory above determines the crossover topology and the scaling of the crossover lengths, but leaves the effective EW amplitude undetermined. For the lowest nonlinearity $n=2$, this amplitude can be estimated self-consistently by combining the crossover form of the structure factor with the Hartree closure. 
Let us now consider the particular value $n=2$, i.,e., the first order correction to linear elasticity, and
proceed as in Sec. \eqref{sec:hartree} to estimate $\tilde\nu_{\mathrm{eff}}$ following the self-consistent Hartree procedure 
\begin{equation}
  \tilde\nu_{\mathrm{eff}}=1 +  \frac {3 c_4}{c_2} \langle(\partial_x h)^2\rangle,
        \label{nu_tilde}
\end{equation}
but instead of adopting the form (\ref{eq:sofqLMtoEW}), as in  Eq. \eqref{eq:integralhartree1}, we now integrate over all wave-vectors, using  Eqs. \eqref{eq:g<} and \eqref{eq:g>} as input
\begin{align}
        \langle(\partial_x h)^2\rangle= &\frac{1}{\pi}\int_0^\infty dq\, q^2 S(q\,|\,T,\tau)=\nonumber \\
        & \frac{3T}{\tilde\nu_{\mathrm{eff}}\pi c_2^{3/2}\sqrt{\tau}}\int_0^\infty dk\,  g(k,\Gamma_4),
        \label{mean_grad}
\end{align}
where the limits of integration are now constant.
Thus, the nondimensional effective elastic constant satisfies the closed equation
\begin{equation}
        \tilde\nu_{\mathrm{eff}}=1 + \frac{G(\Gamma_4)}{\tilde\nu_{\mathrm{eff}}}.
\label{eq:selfconsistenthartree2}
\end{equation}
with solution
\begin{equation}
        \tilde\nu_{\mathrm{eff}}=\frac{1}{2}+\sqrt{\frac{1}{4}+ G(\Gamma_4)}.
\end{equation}
The function $G(\Gamma_4)$ is
\begin{equation}
        G(\Gamma_4)=\frac{3\Gamma_4}{\pi}\int_0^\infty dk\, g(k), 
\end{equation}
which, according to eqs (\ref{eq:g<}) and (\ref{eq:g>}),  also has two branches:

\begin{equation}
G(\Gamma_4)=
\left\{
\begin{array}{lcr}
\dfrac{6}{\pi}\Gamma_4  &\text{if} & \Gamma_4< 2\\
&&\\ 
\dfrac{3}{\pi}\Gamma_4\;[(2\zeta_s-1) \,(\Gamma_4/2)^{-1/5} \\                   + (3-2\zeta_s)\,(\Gamma_4/2)^{(11-12\zeta_s)/5}] &\text{if}& \Gamma_4 > 2
\end{array}
\right.
\end{equation}

Note that when $\tau\to\infty$ ($\Gamma_4\to 0$, almost linear case),
$\tilde\nu_{\mathrm{eff}}\to 1$ ($\nu_{\rm eff}\to c_2$), and the  correlation length at the EW
transition behaves as $\xi \sim l_{\rm LM/EW} \sim \sqrt{c_2\tau}\sim\tau^{1/2}$ as expected from general arguments (Sec. \ref{sec:multiscale}). 
On the other
hand, when $\tau\to 0$ ($\Gamma_4\to\infty$, strongly nonlinear case), $\xi \sim 
l_{\rm ALM/EW} \sim 
\sqrt{\nu_\mathrm{eff}\tau}\sim\tau^{2/5}$ as anticipated in Sections \eqref{sec:multiscale} and \eqref{sec:hartree} with different approaches. Interestingly, the self-consistent equation Eq.~\eqref{eq:selfconsistenthartree2} is different from the one derived previously  employing a more naive procedure that yields Eq.~\eqref{eq:self_cons_naive}. Nevertheless,
the current analysis for $n=2$ show that in this limit $\tilde\nu_{\mathrm{eff}}\sim \Gamma_4^{2/5} \sim \tau^{1/5}$ in agreement with the more naive Hartree approach (see Table \ref{tab:hartreeregimes}).

This framework can be generalized to 
other mixed case,
with only $c_2$ and $c_{2n}$ nonzero, for any $n>1$; which leads to
the crossover diagram sketched in Fig. \ref{fig:crossoverdiagram}. Its advantage is that we can now predict all  crossovers as a function of $\tau$, $T$ and the relative importance of the elastic constants $c_2>0$ and $c_{2n}>0$,  generalizing the results of previous sections and supporting the full crossover diagram of Fig. \ref{fig:crossoverdiagram}.

\section{Numerical Results and Discussion}

 \begin{figure}[t!]
    \centering
    \includegraphics[width=\linewidth]{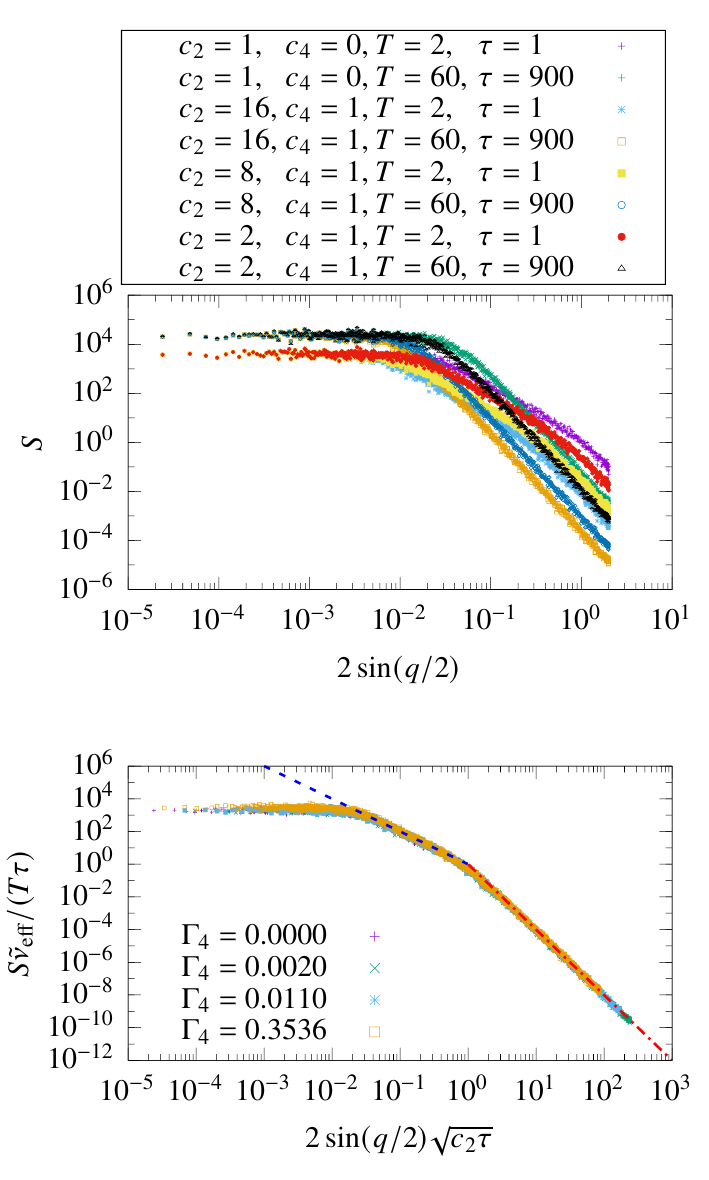}
     \caption{$\Gamma_4<2$. Top: Structure factors for several set of parameters (in the key). Bottom: The same data collapsed onto a  master curve. The line represents the analytical solution
     \color{black}
     $g_<(k,\Gamma_4)/k^2$. 
     \color{black}
     Dashed (blue) part and dash-dotted (red) parts correspond EW and LM regimes, respectively.
    }
    \label{num_gamma_chico}
  \end{figure}

We have combined a phenomenological scaling approach based on the roughening of the ALM model of Ref. \onlinecite{purrello2019roughening} and the self-consistent Hartree method. Since none of these approaches is strictly exact for the model of Eq. \eqref{eq:eqmotion} in this section we perform accurate numerical checks.

\begin{figure}[t!]
    \centering
    \includegraphics[width=\linewidth]{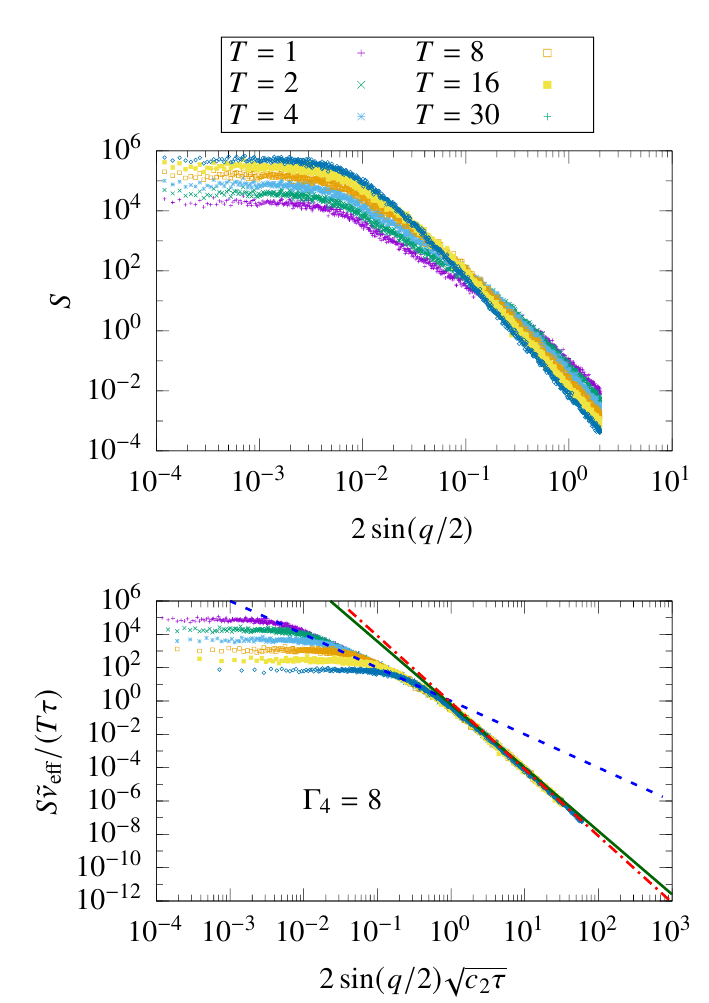}
     \label{gama_8}
    \caption{Structure factors for  $\Gamma_4=8$. The plots correspond to $c_2=1,\, c_4=8$ and several pairs of $T$ (in the key) and $\tau$, which satisfy $T/\sqrt\tau=1$.
    Top panel: raw data. Bottom panel: rescaled data collapsing on an universal function. The line represents the analytical solution
    \color{black}
    $g_>(k,\Gamma_4)/k^2$. 
    \color{black}
    Dashed (blue), solid (dark-green) and dash-dotted (red) parts correspond to EW. ALM and LM regimes, respectively
    } 
    \label{num_gamma_8}
    \end{figure}

We test the analytical predictions by numerical integration of Eq.~\eqref{eq:eqmotion} (see Appendix~\ref{app:numerics}). We compute the structure factor
$S(q,t)=\langle |h(q,t)|^2 \rangle$
starting from a flat initial condition $h(x,0)=0$. At finite times, the system exhibits a time-dependent crossover wavevector $q_{\rm flat}(t)$ separating equilibrated modes ($q>q_{\rm flat}(t)$), for which $S(q,t)\simeq S(q)$, from non-stationary modes ($q<q_{\rm flat}(t)$), which retain memory of the initial condition. All simulations are performed in a regime where $q_{\rm flat}(t)L \gg 1$, ensuring a broad stationary window  ($q>q_{\rm flat}$) for comparison with theoretical predictions.

Our analytical predictions for $\Gamma_4 < 2$ are confirmed in Fig.~\ref{num_gamma_chico}, which illustrates the behavior of the structure factor within this regime.
The top panel displays the raw data obtained through numerical simulations for the various parameter sets indicated in the key, showing distinct curves for each case.
In the bottom panel, \color{black} the stationary part \color{black} of these datasets collapse onto a single master curve, confirming that the scaled behavior becomes independent of $\Gamma_4$.
The solid line represents our analytical solution,
\color{black}
$g_<(k,\Gamma_4)/k^2$ 
\color{black}
(see Eq \eqref{eq:g<}), 
which perfectly captures the numerical data across all scales.
Furthermore, the plot highlights the crossover between two distinct physical behaviors: the dashed (blue) segment marks the Edwards-Wilkinson regime at large length scales (small $k$), whereas the dash-dotted (red) segment traces the transition to the Larkin Model  regime at small length scales (large $k$).

For the regime $\Gamma_4 > 2$, the physical picture changes, 
as the system exhibits a more complex behavior characterized by two distinct crossovers. 
To illustrate this regime, where 
the  master curve for the structure factor becomes explicitly dependent on $\Gamma_4$
Fig.~\ref{num_gamma_8} presents a representative example for the specific case of $\Gamma_4 = 8$.
The top panel displays the raw data obtained through numerical simulations for several pairs of temperature $T$ and characteristic time $\tau$ satisfying $T/\sqrt{\tau} = 1$, as detailed in the key.
In the bottom panel, these datasets successfully collapse onto a single universal function, demonstrating that the scaling holds despite the explicit $\Gamma_4$ dependence.
The solid line represents our analytical solution, 
\color{black}
$g_>(k,\Gamma_4)/k^2$ 
\color{black}
(see Eq. \eqref{eq:g>}), which perfectly captures the numerical results across all wave-vectors.
The sequence of regimes revealed by the two crossovers spans from large to small length scales: the dashed (blue) segment marks the Edwards-Wilkinson (EW) regime at small $k$, the solid (dark-green) segment corresponds to an intermediate ALM regime, and the dash-dotted (red) segment traces the final transition to the Larkin Model (LM) regime at large $k$.

\begin{figure}[t!]
    \centering
    \includegraphics[width=\linewidth]{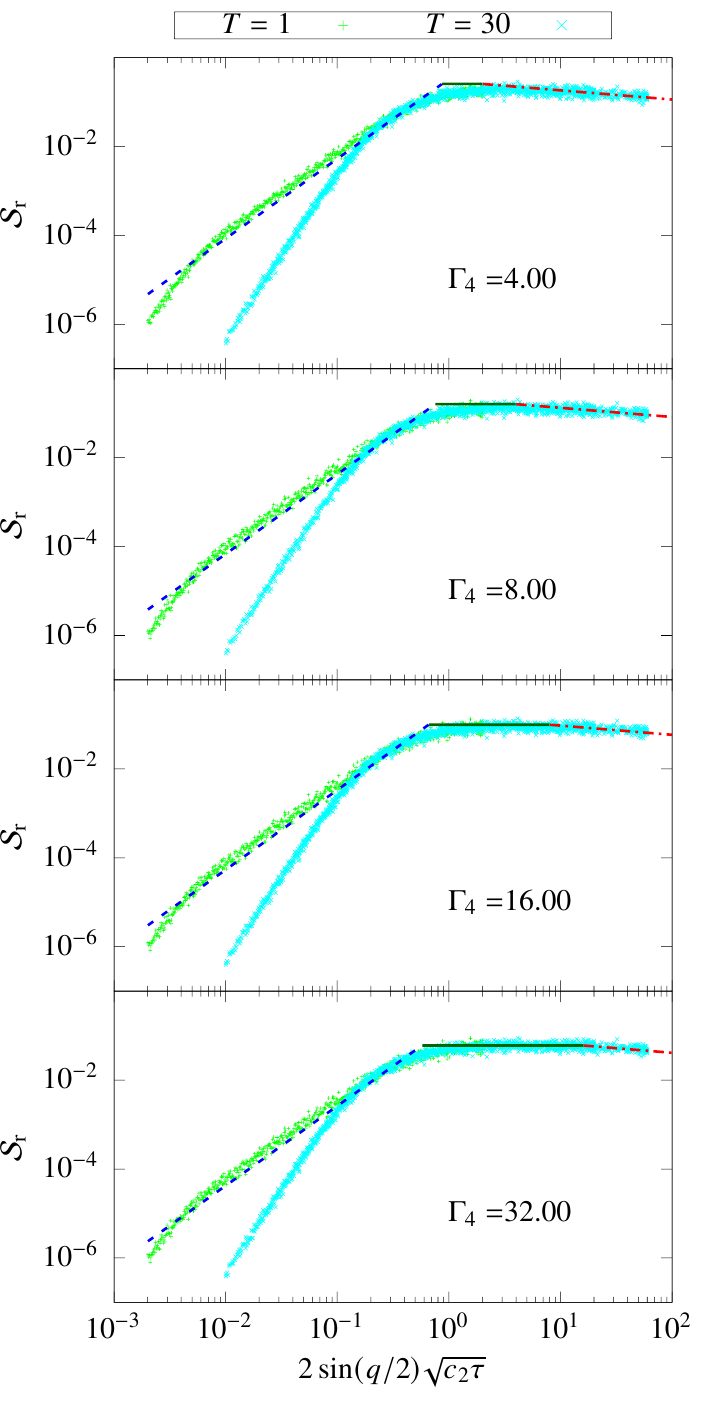}
    \label{S_rot}
       \caption{Structure factor for  $\Gamma_4>2$. The plots correspond to four different $\Gamma_4$ (in the key) and pairs of $T$ (in the key) and $\tau$, which satisfy $T/\sqrt\tau=1$.
    Data points are the results of numerical simulations and the lines represent the analytical solutions: dashed (blue), solid (dark-green) and dash-dotted (red) parts belong to EW. ALM and LM regimes, respectively. In all panels the structure factor is presented in a rotated manner (see eq. (\ref{eq_s_rot})) to better appreciate the widening of the ALM zone with increasing $\Gamma_4$.
    } 
 \end{figure}

To visually confirm how the non-linear elasticity parameter alters the layout of the crossovers, Fig.~\ref{S_rot} presents the structure factor for four different values of $\Gamma_4 > 2$.
In this figure, all datasets are plotted in a rotated manner according to 
\begin{equation}
    \mathcal{S}_\mathrm{r}(q)= \frac{S(q)}{T\tau}\cdot q^{2\zeta_s+1}
    \label{eq_s_rot}
\end{equation}
The data points represent the results obtained through numerical simulations for several pairs of temperature $T$ and characteristic time $\tau$ satisfying $T/\sqrt{\tau} = 1$.
The curves correspond to our analytical solution, where the dashed (blue), solid (dark-green), and dash-dotted (red) segments highlight the EW, ALM, and LM regimes, respectively.
This specific rotated representation is highly effective, as it clearly illustrates that the intermediate ALM zone widens significantly as $\Gamma_4$ grows, as predicted from our analytical framework, 
\color{black} specifically Eq.\eqref{eq:k1andk2} \color{black}.

\section{Conclusions}

We have shown that a nonlinear elastic interface driven by temporally correlated Ornstein--Uhlenbeck noise, which breaks statistical tilt symmetry, exhibits a  stationary crossover structure controlled by a single dimensionless coupling $\Gamma_{2n}$. The resulting phase diagram is organized into three self-affine regimes—Larkin, anharmonic Larkin, and Edwards--Wilkinson—with analytically determined crossover scales.

Combining multiscale scaling arguments, a self-consistent Hartree approximation, and numerical simulations, we obtained a consistent description of the full crossover diagram shown in Fig.~\ref{fig:crossoverdiagram}. The success of the Hartree approximation is consistent with previous observations in elastic interface problems where geometrical observables are accurately captured by Gaussian theories.

Although our analysis focuses on a specific nonlinear-elastic active interface, the underlying mechanism is more general: the competition between persistent fluctuations and nonlinear relaxation generates a hierarchy of scale-dependent roughening regimes whose extent is controlled by a small number of dimensionless parameters. We therefore expect the framework developed here to be useful beyond the particular model studied, providing a simple route to understand finite-size effects, crossover phenomena, and effective large-scale behavior in a broader class of driven and active interfaces.

\begin{acknowledgments}
We thank A. Alés, D. Liarte and A. Rosso for useful discussions.
\end{acknowledgments}

\bibliography{referencias}

@article{Martin2021,
  title = {Statistical mechanics of active Ornstein-Uhlenbeck particles},
  author = {Martin, David and O'Byrne, J\'er\'emy and Cates, Michael E. and Fodor, \'Etienne and Nardini, Cesare and Tailleur, Julien and van Wijland, Fr\'ed\'eric},
  journal = {Phys. Rev. E},
  volume = {103},
  issue = {3},
  pages = {032607},
  numpages = {25},
  year = {2021},
  month = {Mar},
  publisher = {American Physical Society},
  doi = {10.1103/PhysRevE.103.032607},
  url = {https://link.aps.org/doi/10.1103/PhysRevE.103.032607}
}

@article{matsushita1998interface,
  title={Interface growth and pattern formation in bacterial colonies},
  author={Matsushita, Mitsugu and Wakita, J and Itoh, H and Rafols, Ismael and Matsuyama, T and Sakaguchi, H and Mimura, M},
  journal={Physica A: Statistical Mechanics and its Applications},
  volume={249},
  number={1-4},
  pages={517--524},
  year={1998},
  publisher={Elsevier}
}

@article{Bonilla2019,
  title = {Active Ornstein-Uhlenbeck particles},
  author = {Bonilla, L. L.},
  journal = {Phys. Rev. E},
  volume = {100},
  issue = {2},
  pages = {022601},
  numpages = {12},
  year = {2019},
  month = {Aug},
  publisher = {American Physical Society},
  doi = {10.1103/PhysRevE.100.022601},
  url = {https://link.aps.org/doi/10.1103/PhysRevE.100.022601}
}

@article{lemerle1998domain,
  title={Domain wall creep in an Ising ultrathin magnetic film},
  author={Lemerle, S and Ferr{\'e}, J and Chappert, C and Mathet, V and Giamarchi, T and Le Doussal, P},
  journal={Physical review letters},
  volume={80},
  number={4},
  pages={849},
  year={1998},
  publisher={APS}
}

@article{purrello2019roughening,
  title={Roughening of the anharmonic Larkin model},
  author={Purrello, V{\'\i}ctor Hugo and Iguain, Jose Luis and Kolton, Alejandro Benedykt},
  journal={Physical Review E},
  volume={99},
  number={3},
  pages={032105},
  year={2019},
  publisher={APS}
}

@article{metaxas2007creep,
  title={Creep and Flow Regimes of Magnetic Domain-Wall Motion in Ultrathin Pt/Co/Pt Films<? format?> with Perpendicular Anisotropy},
  author={Metaxas, PJ and Jamet, JP and Mougin, A and Cormier, M and Ferr{\'e}, J and Baltz, Vincent and Rodmacq, B and Dieny, B and Stamps, RL},
  journal={Physical review letters},
  volume={99},
  number={21},
  pages={217208},
  year={2007},
  publisher={APS}
}

@article{gorchon2014pinning,
  title={Pinning-dependent field-driven domain wall dynamics and thermal scaling in an ultrathin Pt/Co/Pt magnetic film},
  author={Gorchon, Jon and Bustingorry, Sebastian and Ferr{\'e}, Jacques and Jeudy, Vincent and Kolton, Alejandro Benedykt and Giamarchi, Thierry},
  journal={Physical Review Letters},
  volume={113},
  number={2},
  pages={027205},
  year={2014},
  publisher={APS}
}

@article{marchetti2013hydrodynamics,
  title={Hydrodynamics of soft active matter},
  author={Marchetti, M Cristina and Joanny, Jean-Fran{\c{c}}ois and Ramaswamy, Sriram and Liverpool, Tanniemola B and Prost, Jacques and Rao, Madan and Simha, R Aditi},
  journal={Reviews of modern physics},
  volume={85},
  number={3},
  pages={1143--1189},
  year={2013},
  publisher={APS}
}

@article{bru2003universal,
  title={The universal dynamics of tumor growth},
  author={Br{\'u}, Antonio and Albertos, Sonia and Subiza, Jos{\'e} Luis and Garc{\'\i}a-Asenjo, Jos{\'e} L{\'o}pez and Br{\'u}, Isabel},
  journal={Biophysical journal},
  volume={85},
  number={5},
  pages={2948--2961},
  year={2003},
  publisher={Elsevier}
}

@article{sussman2018soft,
  title={Soft yet sharp interfaces in a vertex model of confluent tissue},
  author={Sussman, Daniel M and Schwarz, JM and Marchetti, M Cristina and Manning, M Lisa},
  journal={Physical review letters},
  volume={120},
  number={5},
  pages={058001},
  year={2018},
  publisher={APS}
}

@article{Cugliandolo_2011,
doi = {10.1088/1751-8113/44/48/483001},
url = {https://doi.org/10.1088/1751-8113/44/48/483001},
year = {2011},
month = {nov},
publisher = {IOP Publishing},
volume = {44},
number = {48},
pages = {483001},
author = {Cugliandolo, Leticia F},
title = {The effective temperature},
journal = {Journal of Physics A: Mathematical and Theoretical}
}

@article{galeano2003dynamical,
  title={Dynamical scaling analysis of plant callus growth},
  author={Galeano, J and Buceta, J and Juarez, K and Pumarino, B and De La Torre, J and Iriondo, JM},
  journal={Europhysics Letters},
  volume={63},
  number={1},
  pages={83},
  year={2003},
  publisher={IOP Publishing}
}

@article{Gompper_2025,
  author    = {Gompper, Gerhard and Stone, Howard A. and Kurzthaler, Christina and Saintillan, David and Peruani, Fernado and Fedosov, Dmitry A. and Auth, Thorsten and Cottin-Bizonne, Cecile and Ybert, Christophe and Clément, Eric and Darnige, Thierry and Lindner, Anke and Goldstein, Raymond E. and Liebchen, Benno and Binysh, Jack and Souslov, Anton and Isa, Lucio and di Leonardo, Roberto and Frangipane, Giacomo and Gu, Hongri and Nelson, Bradley J. and Brauns, Fridtjof and Marchetti, M. Cristina and Cichos, Frank and Heuthe, Veit-Lorenz and Bechinger, Clemens and Korman, Amos and Feinerman, Ofer and Cavagna, Andrea and Giardina, Irene and Jeckel, Hannah and Drescher, Knut},
  title     = {The 2025 motile active matter roadmap},
  journal   = {Journal of Physics: Condensed Matter},
  year      = {2025},
  month     = {feb},
  volume    = {37},
  number    = {14},
  pages     = {143501},
  doi       = {10.1088/1361-648X/adac98},
  url       = {https://doi.org/10.1088/1361-648X/adac98},
  publisher = {IOP Publishing}
}

@article{zagarra2024,
  title = {Infection fronts in randomly varying transmission-rate media},
  author = {Zagarra, Renzo and Laneri, Karina and Kolton, Alejandro B.},
  journal = {Phys. Rev. E},
  volume = {110},
  issue = {3},
  pages = {034308},
  numpages = {9},
  year = {2024},
  month = {Sep},
  publisher = {American Physical Society},
  doi = {10.1103/PhysRevE.110.034308},
  url = {https://link.aps.org/doi/10.1103/PhysRevE.110.034308}
}

@article{KoltonLaneri2019,
  author       = {Kolton, Alejandro B. and Laneri, Karina},
  title        = {Rough infection fronts in a random medium},
  journal      = {European Physical Journal B},
  volume       = {92},
  number       = {6},
  pages        = {126},
  year         = {2019},
  doi          = {10.1140/epjb/e2019-90582-3},
  url          = {https://link.springer.com/article/10.1140/epjb/e2019-90582-3}
}

@book{BarabasiStanley1995,
  author    = {Barab{\'a}si, Albert-L{\'a}szl{\'o} and Stanley, H. Eugene},
  title     = {Fractal Concepts in Surface Growth},
  publisher = {Cambridge University Press},
  address   = {Cambridge},
  year      = {1995}
}

@book{Safran1994,
  author    = {Safran, Samuel A.},
  title     = {Statistical Thermodynamics of Surfaces, Interfaces, and Membranes},
  publisher = {Addison-Wesley},
  address   = {Reading, MA},
  year      = {1994}
}

@book{NelsonPiranWeinberg2004,
  editor    = {Nelson, David R. and Piran, Tsvi and Weinberg, Steven},
  title     = {Statistical Mechanics of Membranes and Surfaces},
  publisher = {World Scientific},
  address   = {Singapore},
  year      = {2004}
}

@book{ChaikinLubensky1995,
  author    = {Chaikin, Paul M. and Lubensky, Tom C.},
  title     = {Principles of Condensed Matter Physics},
  publisher = {Cambridge University Press},
  address   = {Cambridge},
  year      = {1995}
}

@article{Marchetti2013,
  author  = {Marchetti, M. Cristina and Joanny, Jean-Fran{\c{c}}ois and Ramaswamy, Sriram and Liverpool, Tanniemola B. and Prost, Jacques and Rao, Madan and Simha, R. Aditi},
  title   = {Hydrodynamics of soft active matter},
  journal = {Reviews of Modern Physics},
  volume  = {85},
  pages   = {1143--1189},
  year    = {2013},
  doi     = {10.1103/RevModPhys.85.1143}
}

@article{Fodor2016,
  author  = {Fodor, {\'E}tienne and Nardini, Cesare and Cates, Michael E. and Tailleur, Julien and Visco, Paolo and van Wijland, Fr{\'e}d{\'e}ric},
  title   = {How far from equilibrium is active matter?},
  journal = {Physical Review Letters},
  volume  = {117},
  pages   = {038103},
  year    = {2016},
  doi     = {10.1103/PhysRevLett.117.038103}
}

@article{Ales2021,
  title = {Roughening of the anharmonic elastic interface in correlated random media},
  author = {Al\'es, Alejandro and L\'opez, Juan M.},
  journal = {Phys. Rev. E},
  volume = {104},
  issue = {4},
  pages = {044108},
  numpages = {7},
  year = {2021},
  month = {Oct},
  publisher = {American Physical Society},
  doi = {10.1103/PhysRevE.104.044108},
  url = {https://link.aps.org/doi/10.1103/PhysRevE.104.044108}
}

@article{Rosso2003,
  title = {Universal interface width distributions at the depinning threshold},
  author = {Rosso, Alberto and Krauth, Werner and Doussal, Pierre Le and Vannimenus, Jean and Wiese, Kay J\"org},
  journal = {Phys. Rev. E},
  volume = {68},
  issue = {3},
  pages = {036128},
  numpages = {4},
  year = {2003},
  month = {Sep},
  publisher = {American Physical Society},
  doi = {10.1103/PhysRevE.68.036128},
  url = {https://link.aps.org/doi/10.1103/PhysRevE.68.036128}
}

@article{Moulinet2004,
  title = {Width distribution of contact lines on a disordered substrate},
  author = {Moulinet, S\'ebastien and Rosso, Alberto and Krauth, Werner and Rolley, Etienne},
  journal = {Phys. Rev. E},
  volume = {69},
  issue = {3},
  pages = {035103(R)},
  numpages = {4},
  year = {2004},
  month = {Mar},
  publisher = {American Physical Society},
  doi = {10.1103/PhysRevE.69.035103},
  url = {https://link.aps.org/doi/10.1103/PhysRevE.69.035103}
}

@article{ledoussal2003,
  title = {Higher correlations, universal distributions, and finite size scaling in the field theory of depinning},
  author = {Le Doussal, Pierre and Wiese, Kay J\"org},
  journal = {Phys. Rev. E},
  volume = {68},
  issue = {4},
  pages = {046118},
  numpages = {15},
  year = {2003},
  month = {Oct},
  publisher = {American Physical Society},
  doi = {10.1103/PhysRevE.68.046118},
  url = {https://link.aps.org/doi/10.1103/PhysRevE.68.046118}
}

@misc{droykttonActiveInterface,
  author       = {droyktton},
  title        = {{ActiveInterface}},
  year         = {2026},
  publisher    = {GitHub},
  journal      = {GitHub repository},
  howpublished = {\url{https://github.com/droyktton/ActiveInterface/tree/main}},
  note         = {Accessed: June 8, 2026}
}


\appendix
\section{Numerical Simulation}
\label{app:numerics}
To solve Eq.~\eqref{eq:eqmotion} numerically we discretize space as
\begin{align}
\partial_t h_i &= \frac{c_2}{dx^2} (h_{i+1} + h_{i-1} - 2h_i) \nonumber \\
&+ \frac{c_{2n}}{dx^{2n}} \big[(h_{i+1} - h_i)^{2n-1} - (h_i - h_{i-1})^{2n-1}\big] + \eta_i(t),
\label{eq:eqmotiondiscrete}
\end{align}
with $i = 0, \dots, N-1$, using periodic boundary conditions for an interface of length $L$, i.e. $h_0 = h_{N-1}$, and $N = L/dx$ grid points.

The colored noise $\eta_i(t)$ is generated from
\begin{align}
\tau \partial_t \eta_i = -{\eta_i(t)} + \xi_i(t),
\label{eq:noiseeq}
\end{align}
where $\xi_i(t)$ is a Gaussian uncorrelated white noise satisfying
$\langle \xi \rangle = 0$ and
$\langle \xi_i(t) \xi_j(t') \rangle = 2T \delta_{ij} \delta(t - t')$.
The steady-state solution of Eq.~\eqref{eq:noiseeq} satisfies the desired correlations of Eq.~\eqref{eq:colorednoise}.
Accordingly, all simulations of Eqs.~\eqref{eq:eqmotiondiscrete} and \eqref{eq:noiseeq} are initialized with a steady-state realization of $\eta_i(0)$, and the initial condition $h_i(0)=0$.
The case of a long-range temporal correlated noise was analyzed in Ref.~\onlinecite{Ales2021}.

Time integration of Eqs.~\eqref{eq:eqmotiondiscrete} and \eqref{eq:noiseeq} is performed using explicit methods with appropriate spatial and time discretizations, $dx$ and $dt$ respectively. The scheme can be efficiently implemented for large $N$ using parallel GPGPU computing. The CUDA/C++ code is freely available in Ref. \onlinecite{droykttonActiveInterface}.

\end{document}